# The restrictions of the Maximum Entropy Production Principle

L. M. Martyushev[1,2], V.D. Seleznev[2]

[1] Institute of Industrial Ecology, Russian Academy of Sciences, 20 S. Kovalevskaya St. Ekaterinburg, 620219 Russia;

[2] Ural Federal University, 19 Mira St., Ekaterinburg 620002, Russia

e-mail: LeonidMartyushev@gmail.com

This mini review is the first study where we shall structure, list and briefly discuss the existing restrictions of the maximum entropy production principle (MEPP) in order to explain typical misconceptions existing among MEPP critics.

The maximum entropy production principle (MEPP) is known in the literature for a long time and became a useful tool for solving various problems of physics, environmental science, biology, etc. The information about the history of the principle, the formulations, kinetic and thermodynamic grounds of MEPP, as well as the modern applications can be found in the reviews [1-6]. However, in spite of these studies, the modern literature lacks a unified and clear summary of MEPP restrictions. As a consequence, this gave and still gives rise to the publications which attempt to disprove the principle using a number of examples that are outside the range of the principle's applicability [7-12]. A recent critical publication by J.Ross et al. [8] confirms this fact. A number of questions brought up in this and similar papers are very important and useful for discussion. Herein, we will try to explain a number of fundamental points connected with MEPP, more precisely define the conditions of its applicability, and show the groundlessness of the published contrary instances.

The thermodynamic formulation of the classical MEPP was proposed by H. Ziegler [13-15]. Here is one of its variants: if the thermodynamic forces $X_i$ are preset, then the true thermodynamic fluxes $J_i$ satisfying the side condition $\sigma = \sum_i X_i J_i$ give the maximum value of the entropy production density $\sigma(\mathbf{J})$. In fact, this local principle is one of the simplest variational formulations of classical nonequilibirum thermodynamics. According to the classical formulation of MEPP, for the kinetic level of description, the entropy production functional is varied by distribution function rather than by fluxes [3].

If a dependence of entropy production on fluxes $\sigma(\mathbf{J})$ (in the thermodynamic version) or on particle distribution (in the kinetic version) is defined, then MEPP can be used to determine the explicit relationship between thermodynamic forces and fluxes $X_i(\mathbf{J})$. So, according to [3, 13-15], it follows from the above formulation that:

$$X_i = \frac{\sigma(J)}{\sum_i (J_i \cdot \partial\sigma/\partial J_i)} \partial\sigma/\partial J_i \qquad (1)$$

As can be seen from Eq.(1), a relationship between thermodynamic fluxes and forces can be both linear and *nonlinear*. This is an important corollary of Ziegler's principle. For the simplest case when $\sigma = \sum_{i,k} R_{ik} J_i J_k$ ($R_{ik}$ is the constant coefficient matrix), the main equations of linear nonequilibrium thermodynamics:

$$J_i = \sum_k L_{ik} X_k \qquad (2)$$

and the reciprocal relation for the kinetic coefficients $L_{ik}$ ($L_{ik} = L_{ki}$) can be easily obtained from Eq.(1) [3, 13-15]. Thus, MEPP enables to deductively (rather than by generalizing experimental data) obtain the equations of Fourier, Ohm, Navier, Fick et al., as well as cross connections between fluxes and forces. If, for example, a fourth-degree polynomial instead of a quadratic polynomial is selected as $\sigma(J)$, the nonlinear relationship between thermodynamic fluxes and forces used for the study of plasticity and viscoelastic systems can be obtained [13-15].

The obtained explicit dependences $X_i(J)$ together with the conservation laws allow writing the equations of heat-and-mass transfer, momentum (the Navier-Stokes equation), charge (the Kirchhoff laws), chemical kinetics, etc. in the closed form. These equations (mostly differential) completed with boundary and initial conditions are used to describe various processes observed in nature. MEPP makes no conclusions regarding properties of the solutions to these equations! Entropy production can demonstrate here any behavior. Thus, only a relationship between fluxes and forces can be found from the maximum entropy production, only in this case Ziegler's MEPP is true. Let us call it the first restriction of the principle.

Zigler's MEPP is a variational principle of classical irreversible thermodynamics. The local equilibrium hypothesis is the main hypothesis of classical irreversible thermodynamics [16-19]. According to it (see, e.g. [18]): "the local and instantaneous relations between thermodynamic quantities in a system out of equilibrium are the same as for a uniform system in equilibrium." This local equilibrium hypothesis limits the applicability range of classical irreversible thermodynamics and, therefore, the applicability range of MEPP (the second restriction of the principle)[1].

As is well known, there is a number of statements logically connected with the local equilibrium hypothesis (the second restriction of MEPP) [16-19]. Let is recall the main ones: (1) A system under study can be presented as a set of cells that are big enough to be considered as macroscopic thermodynamic subsystems (i.e. the number of particles and their collisions in a local element of the volume should be very large) but small enough to reach equilibirum within a period much shorter than the system relaxation time; (2) $\sigma$ is not negative for any values of variables (in particular, thermodynamic fluxes[2]).

Thus, according to the first two restrictions, MEPP allows finding only a local relationship between thermodynamic fluxes and forces. In other words, the extremized entropy production should not represent a space/time integral[3]. However, a question arises: why maximization of the total (global, integral) entropy production composed of many local areas shows good results in a number of cases (the studies by G. Paltridge [20] on the calculation of the Earth's climate are the most well known)? This is a very complicated question and we do not have the ultimate answer yet. The possible answer can be as follows. As is known, a sum of particular solutions is also a solution in the case of *linear* problems. By analogy, the following hypothesis can be advanced for *linear* systems: the local relationship $X_i(J)$ found from Ziegler's local MEPP corresponds to the fluxes and forces following from the maximization of integral entropy production, i.e. the local maximization of entropy production for a *linear* system is equivalent to the maximization of integral entropy production. The discussion and reasoning of this statement as well as the necessary conditions for its validity are given in the paper [21]. Based on the above, it seems that Paltridge's success is explained by the fact that the relationship between fluxes and forces in his climatic problem is close to *linear* (or that $\sigma(J)$ is close to a second-degree polynomial). Consequently, Paltridge's integral maximization of entropy production reduces to Ziegler's local one. In the general case (if there is doubt whether a relationship between fluxes and forces is close to linear or whether $\sigma(J)$ has little difference from a second-degree polynomial), the integral maximization of entropy production may cause errors.

---

[1] A comparison of the corollaries of classical irreversible thermodynamics with the experiment showed that the local equilibrium hypothesis proves itself very well for the majority of nonequilibrium states that we usually see in nature as well as for the majority of scales (from microscopic to galactic). This hypothesis is invalid, for instance, when shock waves are considered.

[2] The critics of MEPP often forget about this obvious corollary of the second law [12].

[3] In this connection, the "disproving" example on the page 7859 (see Eq.(1),(2) and Fig.1) of Ref.[8] is absolutely inappropriate.

In addition to the mentioned restriction, there are other restrictions that are explicitly or implicitly specified in the fundamental papers by Ziegler [13-15] (as well as in the review [3]). Let us enumerate them:

The third restriction. It is assumed that the entropy production is a known function of fluxes $J$, and the dependence $\sigma(J)$ is not constant, i.e. the maximum value of $\sigma$ can be **chosen** when the $J$ changes.

The fourth restriction. Let us assume that there are $n$ nonequilibrium processes simultaneously occurring in the system, and let us designate their thermodynamic fluxes as $J_1,\ldots, J_n$. According to H. Ziegler (see, for example, item 14.4 [14]), such nonequilibrium processes can always be divided into two types: compound and complex. In the case of compound processes, the entropy production of a process at hand can be represented as a sum of functions dependent only on some (but not on all) of the fluxes. In particular, a compound process can be divided on two subprocesses $\sigma(J_1,\ldots,J_n)=\sigma_1(J_1,\ldots,J_k)+\sigma_2(J_{k+1},\ldots,J_n)$, where $\sigma_1$ and $\sigma_2$ are two functions dependent on different fluxes. There may be more subprocesses but it is important that the fluxes determining one of the subprocesses are not included into (do not influence) the other ones. Simultaneous chemical and thermal processes can be the simplest example of compound processes: the former are scalar and the latter are vector; and according to Curie's principle, these processes cannot influence each other. If a nonequilibrium process cannot be represented as a sum of entropy productions of individual nonequilibrium subprocesses, then such a process is referred to as complex. In other words, due to the mutual influence (cross phenomena, if we use the terms of nonequilibrium thermodynamics), the entropy production of an aggregate system (process) is not an additive function of entropy productions of subsystems (subprocesses). For example, the entropy production of a complex process composed of two subprocesses, as is well known for linear nonequilibrium thermodynamics, has the form[4] $J_1^2+J_2^2+2J_1J_2$, where the last summand determines nonadditivity (it makes the system complex). A process where the diffusion and heat conduction simultaneously occur is an example of a complex process. As Ziegler points out, MEPP is valid for complex systems and invalid for compound systems. This is a very important restriction, which is often forgotten when discussing Ziegler's principle.

The fifth restriction. A one-one correspondence between the thermodynamic forces and fluxes is assumed. In his classical studies, Ziegler more than once emphasises the importance of this restriction. It is very strong and does not allow describing, in particular, nonequilibrium transitions (for example, a transition from laminar to turbulent flow). This was one of the reasons for formulation of a generalized MEPP (see below).

Let us note that, for the case of *linear* nonequilibrium thermodynamics, Ziegler's MEPP can be considered as a rigorously mathematically proven statement confirmed by numerous experiments and microscopic examination within the scope of the Boltzmann kinetic equation rather than a principle (for more detail, see Ref. [3])[5]. It can be easily seen that all the "disproving" examples given in the papers [7-12] fail to satisfy either the principle itself or some of the restrictions of Ziegler's MEPP.

In addition to the classical MEPP by Ziegler (and its kinetic version), the modern literature has different generalizations of it. Based on the analysis of entropy production behavior in different systems, we proposed the following formulation [5, 22]: at each hierarchical level of the evolution, with preset external constraints, a local relationship between the cause and the response of a nonequilibrium system is established in order to maximize the local entropy production. This formulation of MEPP primarily focuses on expanding the range of applicability of the classical principle: it considers the spontaneous evolution of complex physical, chemical, biological and similar systems for which the use of classical local nonequilibrium thermodynamics only is insufficient in order to predict their behavior. Let us illustrate it with an example[6]: crystallization from a solution. The fluxes of matter and heat as functions of gradients of chemical potential and temperature are established in this system according to Ziegler's principle. However, this is only one of the hierarchical levels of process development. In the case of nonequilibrium growth, the crystal loses its morphological stability upon reaching certain size. As a result, the crystal shape changes considerably (for example, a transition from a regular polyhedron to a dendrite occurs). Such a transition results from a supersaturation in the solution (cause), and the system responses by complicating the shape. This bifurcational transformation, as is shown in Refs.[23,24], occurs with the maximization of the local entropy production. Thus, this is another level of evolution where the generalized MEPP is true.

---

[4] For simplicity, the kinetic coefficients are taken as 1.
[5] For nonlinear nonequilibrium thermodynamics, Ziegler's MEPP can be presently considered as a working hypothesis.

The generalization of Ziegler's MEPP does not imply cancellation of the above restrictions of the applicability range[7]. As follows from the papers [7-11], the biggest misunderstanding relates to the fourth restriction. Let us dwell on it again. The notion of "complex system" means much more than the emphasis of the fact that the system is far from simple. As is known [25-27], a system consisting of a large number of dissimilar elements with nonlinear relations takes on new, sometimes very unexpected properties, that are difficult (and maybe even impossible) to deduce from the study of individual elements or relations. This property is called "emergence" and represents the key feature of a truly complex system. It is appropriate to quote Aristotle here: "the whole is greater than the sum of its parts". So, the properties of a large collective of molecules (which number equals, for example, the Avogadro number) are different from the sum of the properties of its constituent molecules, and the properties of a biological population are not deducible only to the properties of individuals. From the standpoint of entropy production (as well as entropy itself), the complexity of a system (process) is demonstrated by nonadditivity of this quantity when the system (process) is formed from its constituent subsystems (subprocesses). Let us use the examples from chemical kinetics (the Schlögl model and the like) that are often cited to disprove the generalized MEPP [7-10]. Such models represent a small set of consecutive or parallel chemical reactions that **have no effect** on each other. If the properties of individual reactions (specifically, their rate coefficients) are known, then the behavior of a chemical process they constitute can be fully understood (calculated); the entropy productions of individual reactions fully determine the entropy production of the whole process (since the total entropy production is a simple sum of the entropy productions of individual processes). Obviously, in view of the above, the system is not complex. Therefore, the behavior of entropy production for such a compound (not complex) system may fail to conform with MEPP. The generalized MEPP is inapplicable to such systems. Moreover, it is unlikely that some common regularity or principle could be found for such systems (due to their completely deterministic nature). In this regard, we fully support the conclusions of R. Landauer [28, 29]. If properties of elements are known and remain invariable[8] when the elements are combined into a system, any experienced chemist or radio engineer can design a compound system of elements such that the system's entropy production or any other quantity would behave in any predefined way when chemical or electrical potentials (thermodynamics forces) change. Let us state again that the generalized MEPP is valid for complex systems (at least now, there are no refutations of this fact in the literature); moreover, we believe that this principle will soon become a *basic* principle for studying complex emergent systems of different nature (including informational, biological, economical, and social).

Ziegler's classical principle, as is well known [13-15], describes the maximization of entropy production when varying by fluxes with preset forces and vice versa (i.e., this is a symmetrical statement relative to fluxes and forces). For the generalized principle formulated above, the cause can be a generalized analog of both a thermodynamic flux and a force; at the same time, the response of a system can be a generalized analog of both a thermodynamic force and a flux, correspondingly. Today, there is an open question whether the symmetry between fluxes and forces is disturbed during the maximization of entropy production in the general case. A number of studies point at the fact that the entropy production *minimizes* when fluxes are replaced with forces for some hydrodynamic problems [1, 30-34]. However, these studies cannot be used as contrary instances to MEPP for two reasons: (1) the nonlocality of problem formulation (series-connection and parallel-connection systems), and the setting (rather than finding from the maximization) of a relationship between fluxes and forces [30,31]. (2) In the case of a transition from laminar to turbulent flow [32-34], entropy productions in these two nonequilibrium phases are compared using empirical dependences between pressure drop and Reynolds number. The entropy productions of laminar and turbulent flows of an incompressible fluid should be compared for the same volumes with the same kinetic energy (it is appropriate here to recall the following analogy with the Second Law of thermodynamics: a system with the maximum entropy is found only among isolated systems with the same volume, energy, and number of particles). In this case, for a fixed Reynolds number (flow), the maximum entropy production is realized (a thermodynamic force is maximized) [32-34]; whereas for a fixed pressure drop (thermodynamic force), both turbulent and laminar flows have the same entropy production (as calculated for the same kinetic energy of a moving incompressible flow). As follows from the above, the authors [34] have compared the entropy productions incorrectly.

---

[6] Other examples can be found in the papers [3,5,6].
[7] The fifth restriction is assumed redundant. Apparently, additional restrictions or, on the contrary, reduction of the restrictions when formulating the generalized MEPP is possible; however, this requires additional study.
[8] I.e., these elements do not influence each other in an integral system.

Another misconception that we would like to briefly discuss here can be characterized by a typical quotation from Ref. [8]: "for deterministic kinetic systems there is no need, indeed no place, for any additional principle" (see also a similar phrase in the abstract [8]). Let us give two counter-arguments to illustrate the invalidity of this conclusion. (1) Classical mechanics forming the very foundation of determinism has been developing various local and integral variational principles for centuries as an alternative and an addition to the classical Newton approach. Thanks to the work of the best mathematicians and mechanicians such as P. Fermat, L. Euler, J. Lagrange, C. Gauss, W. Hamilton, etc., a number of principles became the part of mechanics. The principle of least action was among them, which subsequently outgrew its mechanical origin and became a basis and a source of progress for a number of branches of theoretical physics. (2) Presently, there are rather simple deterministic nonlinear systems with extremely complex behavior (the so-called deterministic chaos). Using traditional analytic and numerical methods, the evolution of such systems (e.g., some hydrodynamic ones) can be predicted only on a very small time interval. One of the possible methods to analyze such systems is to consider them as being partially chaotic and to apply statistic methods (including those using entropy and MEPP) (see e.g. [35]). (3) The other counter-argument roots in the foundations of mathematics itself, i.e. in the basic language used to describe and analyze almost any natural-science problem (including a deterministic one). This is a well-known theorem of incompleteness by K. Gödel. According to this theorem, any sufficiently complex consistent theory has a statement in it that can be neither proved nor disproved within this theory. It appears that this statement (principle) can be added to the theory at hand without breaking its consistency [36, 37].

The question about the "scientific character" of MEPP touched upon in the critical paper [8] is the last important point we would like to address here. This question was repeatedly raised in the literature and is mainly considered from the well-known viewpoint of K. Popper's falsifiability. A number of authors [38, 39] support the opinion that MEPP (most commonly its generalized version is meant) is only a statistical procedure for obtaining the best prediction of nonequilibrium system behavior in the conditions of insufficient information about the system. As a consequence, they consider the principle to be unfalsifiable. We disagree with that conclusion because there are multiple ways of MEPP falsification which can be proposed according to Popper. The measurement of entropy production at the moment of bifurcation of a complex system (which conforms with the restrictions 1-4) can serve as the simplest experiment for disproving the generalized MEPP. If a state with the minimum entropy production is the most probable realizable state, then the generalized MEPP according to Popper will be disproved. Therefore, in the context of the conception of Popper and a number of his followers, such a principle cannot be called unscientific. We believe, however, that the process of disproving any principle, not only MEPP, is more difficult than it may seem at first sight[9]. As a consequence, it is too simple to consider this problem only from K. Popper's position, especially after the critical publications on the methodology of scientific research programs by I. Lakatos (1970-1978). In this connection, following the philosophic reasoning by Lakatos, the increment of actual knowledge based on the predicting power of a method is the main criteria of the *scientific* method. The recent avalanche-like growth of the number of papers related to MEPP that show interesting and experimentally verified results in different fields of science – from physics to biology – are the best evidences of both scientific character and importance of the Maximum Entropy Production Principle.

Acknowledgments. The authors would like to thank the referees, including Ralph Lorenz, for criticism and a number of interesting questions that helped to improve this paper.

---

[9] Here, it is appropriate to cite H. Poincaré (see Ch. VIII in *Science and Hypothesis*; The Walter Scott Publishing Co., 1904) on the impossibility to falsify the energy conservation law.